\documentclass[10pt, notitlepage, twocolumn, superscriptaddress, nofootinbib, floatfix]{revtex4-1}

\usepackage{amsmath}
\usepackage{mathrsfs}
\usepackage{multirow}
\usepackage{amssymb}
\usepackage{mathtools}
\usepackage{hyperref}
\usepackage[usenames,dvipsnames]{xcolor}
\hypersetup{colorlinks=true, citecolor=Purple, linkcolor=Purple,
urlcolor=Purple}

\usepackage{graphicx}
\usepackage{orcidlink}

\newcommand{\GB}{\mathscr{G}}

\begin{document}

\allowdisplaybreaks

\title{Quasinormal modes of hairy black holes in shift-symmetric theories}

\author{\textbf{Georgios~Antoniou}}
\email{georgios.antoniou@roma1.infn.it}
\affiliation{Dipartimento di Fisica, ``Sapienza'' Universit\'a di Roma \& Sezione INFN Roma1, P.A. Moro 5, 00185, Roma, Italy}
\affiliation{Dipartimento di Fisica,  Universit\`a di Pisa, 56127 Pisa, Italy}

\begin{abstract}
    Alternative theories of gravity may be put to the test by making use of gravitational wave observations. Scalar-tensor theories provide a relatively simple framework that allows for compact object solutions with characteristics different from those of their general relativity counterparts. In shift-symmetric theories, black hole hair may arise when a linear coupling with the Gauss-Bonnet (GB) invariant is introduced.
    The effect of this coupling on the black hole quasinormal modes (QNMs) has been shown to be small, therefore limiting the observational interest in the theory.
    In general, however, one expects that additional shift-symmetric terms in the Lagrangian may be relevant, and indeed, it has been shown that they can have a nontrivial impact on the mass and scalar charge of the stationary solutions. In this work we explore the effect that various such terms have on the axial and polar QNMs of the hairy black hole solutions.
\end{abstract}

   \maketitle

\section{Introduction}
The recent developments in the field of gravitational wave astronomy \cite{LIGOScientific:2016aoc,LIGOScientific:2018mvr,LIGOScientific:2020ibl,LIGOScientific:2021usb,KAGRA:2021vkt} have opened a never-accessed before window onto the strong-field nature of gravitation.
The study of merging compact objects can potentially allow us to experimentally confront the validity of general relativity (GR) against other candidate theories.
In GR black holes may be fully described by their mass, spin, and electric charge \cite{Hawking:1971vc,Carter:1971zc}, and no additional Standard Model parameters are expected \cite{Bekenstein:1972ny,Teitelboim:1972qx} to be relevant.
Observational deviations from the picture predicted by GR could indicate the presence of modifications or a breakdown of the theory. That is why current and future GW observations will be seminal in the search of new fundamental physics \cite{Barack:2018yly,Sathyaprakash:2019yqt,Barausse:2020rsu,Kalogera:2021bya,LISA:2022kgy}.

A simple way to modify GR is by introducing a scalar field in addition to the metric tensor. The Horndeski theory provides the most general framework that in four dimensions describes gravity through a metric tensor supplemented by a scalar field \cite{Horndeski:1974wa,Deffayet:2009mn,Kobayashi:2011nu,Clifton:2011jh}.
In this framework, many works have been presented over recent years, exploring solutions of black holes (and other compact objects) that evade no-hair theorems \cite{Bekenstein:1995un,Hawking:1972qk,Sotiriou:2011dz} and carry secondary scalar hair \cite{Mignemi:1992pm,Kanti:1995vq,Sotiriou:2013qea,Sotiriou:2014pfa,Antoniou:2017acq,Silva:2017uqg,Doneva:2017bvd,Antoniou:2017acq}.
This is achieved through a particular combination of the arbitrary functions appearing in the Horndeski Lagrangian, that eventually may be re-written as a non-trivial coupling of the scalar field with the Gauss-Bonnet (GB) invariant. This particular coupling allows for the evasion no-hair theorems postulated for scalar-tensor gravity \cite{Hui:2012qt}.

In general, massless or almost massless scalars are expected to be observationally favourable. Therefore, shift-symmetric scalars are of high interest as shift symmetry prevents the scalar field from acquiring mass.
As has been discussed in \cite{Saravani:2019xwx}, the theories derived from the Horndeski Lagrangian can be divided into three separate classes based on the conditions under which GR can be a solution of the theory. The first class corresponds to theories that admit GR in any background, the second class to theories admitting GR only in the Minkowski limit, while the third one contains all remaining cases.
It was also shown that all of the second class theories may be expressed as a first class theory plus the shift-symmetric coupling of the scalar field with the GB invariant.
In \cite{Thaalba:2022bnt} a general shift-symmetric Lagrangian was studied were the GB coupling was supplemented with other leading or next-to-leading order shift-symmetric terms. It was shown that despite the fact that the term directly responsible for the existence of the scalar hair was the GB one, the additional terms contributed to the properties and existence conditions for the hairy black hole solutions.
Specifically, the additional terms influence the existence conditions for hairy black holes, and consequently affect the black hole minimum size, which however can never be arbitrarily decreased. The additional terms can also impact the scalar charge that the black holes carry for a certain mass, especially for lighter black holes.

Returning to the discussion regarding the observability of beyond GR effects, one may stress out the importance of the examination of the final stage of a black hole merger, namely the ringdown phase, through the study of the black hole quasinormal modes (QNMs).
The latter correspond to the exponentially dumped sinusoids dominating this final stage of the black hole binary, and their study may allow for an observational test of GR in the strong-field gravitational regime.
QNMs have been studied in the context of theories beyond GR involving a scalar field.
The authors of \cite{Blazquez-Salcedo:2016enn,Blazquez-Salcedo:2017txk} have studied them for spherically symmetric black holes in the Einstein-dilaton-Gauss-Bonnet (EdGB) theory, while more recently QNMs have been explored in theories allowing for \textit{spontaneously scalarized} solutions \cite{Blazquez-Salcedo:2020rhf,Blazquez-Salcedo:2020caw}. In both of these general scenarios the QNMs of the hairy solutions have been shown to be very close to those of their GR counterparts.
More recently, it was demonstrated that other terms which are not directly associated with sourcing the scalar hair -but are important when one considers issues such as the viability of the solutions \cite{Antoniou:2020nax,Antoniou:2021zoy,Ventagli:2021ubn}- have a strong effect on the QNMs of the hairy solutions \cite{Antoniou:2024gdf}.

Here we explore the effects of additional shift-symmetric terms on the QNM spectrum of the black holes.
In Sec.~\ref{sec:background} we summarize the basic formulation of Horndeski gravity, we present the theory which we will study, and we revisit the stationary background solutions.
In Sec.~\ref{sec:perturbations} we introduce the metric and scalar perturbations and we separate our analysis into the axial and polar sector.
In Sec.~\ref{sec:astrophysics} we discuss what the astrophysical implications of the current study could be and finally, in Sec.~\ref{sec:conclusions} we present our conclusions and provide some future perspectives.

\section{Framework and background solutions}
\label{sec:background}

\subsection{Shift-symmetric Horndeski gravity}
As metioned earlier, Horndeski's theory is the most general four-dimensional diffeomorphism-invariant theory involving a metric tensor and a scalar field that leads to second-order field equations upon variation \cite{Horndeski:1974wa,Deffayet:2009mn}. As we also explained, we will restrict our analysis to shift symmetric theories. The shift-symmetric Horndeski action is then given by:
\begin{equation}
\label{eq:Horndeski}
S=\frac{1}{2k}\sum_{i=2}^{5}\int \mathrm{d}^4x\,\sqrt{-g}\mathcal{L}_i+S_\text{M},
\end{equation}
where each sub-Lagrangian $\mathcal{L}_i$ is given in terms of the arbitrary functions $G_i$ which depend on the kinetic term $X=-\nabla_\mu\phi\nabla^\mu\phi/2$. Specifically:
\begin{align}
\mathcal{L}_2 = & \, G_2(X)\,,\\
\mathcal{L}_3 = & \, -G_3(X)\Box\phi\,,\\
\mathcal{L}_4 = & \, G_4(X)R+G_{4X}[(\Box\phi)^2-(\nabla_\mu\nabla_\nu\phi)^2]\,,\\
\mathcal{L}_5 = & \, G_5(X)G_{\mu\nu}\nabla^\mu\nabla^\nu\phi
- \frac{G_{5X}}{6}\big[\left(\Box\phi\right)^3\\
&-3\Box\phi(\nabla_\mu\nabla_\nu\phi)^2+2(\nabla_\mu\nabla_\nu\phi)^3\big]\,,
\end{align}
$R$ is the Ricci scalar, and $G_{\mu\nu}$ is the Einstein tensor. We have also defined $k=8\pi G/c^4$ with $S_\text{M}$ being the matter action. Unless stated otherwise, we will be considering a notation where we set the physical constants to one, \textit{i.e.} $G=c=1$.
Matter is assumed to only couple minimally to the metric, that is, we are working in the so-called Jordan frame.
For the theory described by \eqref{eq:Horndeski}, there exists a conserved Noether current associated with the symmetry \cite{Sotiriou:2014pfa}, namely:
\begin{widetext}
\begin{align}
        &
        J^\mu =-\partial^\mu\phi \big( G_{2X} - G_{3X} \Box \phi +G_{4X} {R} + G_{4XX} \big[ (\Box \phi)^2-(\nabla_\rho\nabla_\sigma\phi)^2 \big]+G_{5X}G^{\rho\sigma}\nabla_{\rho}\nabla_{\sigma}\phi-\frac{G_{5XX}}{6} \big[ (\Box \phi)^3
        \nonumber\\
        &
        -3\Box \phi(\nabla_\rho\nabla_\sigma\phi)^2 + 2(\nabla_\rho\nabla_\sigma\phi)^3 \big] \big)+\partial^\nu X \big(\delta^\mu_\nu G_{3X}-2 G_{4XX} (\Box \phi \delta^\mu_\nu-\nabla^\mu\nabla_\nu \phi)-G_{5X} G^\mu{}_\nu\\
        &
        +\frac{1}{2} G_{5XX} \big[ \delta^{\mu}_{\nu}(\Box\phi)^2- \delta^{\mu}_{\nu}(\nabla_\rho\nabla_\sigma\phi)^2 -2\Box\phi \nabla^\mu\nabla_\nu\phi+2\nabla^\mu \nabla_\rho \phi \nabla^\rho \nabla_\nu \phi \big] \big)+2G_{4X} {R}^{\mu}{}_{\rho} \nabla^\rho \phi \nonumber\\
        &+ G_{5X} \big( -\Box \phi {R}^\mu{}_\rho \nabla^\rho\phi + {R}_{\rho\nu}{}^{\sigma\mu} \nabla^\rho\nabla_\sigma\phi \nabla^\nu\phi+{R}_\rho{}^\sigma \nabla^\rho\phi  \nabla^\mu\nabla_\sigma\phi \big)\,. \nonumber
\end{align}
\end{widetext}
By varying the shift-symmetric Lagrangian with respect to the metric tensor and the scalar field we may derive the Einstein and scalar field equations respectively:
\begin{equation}
    G_{\mu\nu}=T^{(\phi)}_{\mu\nu}\quad , \quad  \nabla^\nu \phi= J^\nu\, .
\end{equation}
For the background geometry we consider a spherically symmetric and static spacetime described by the following metric element:
\begin{equation}
    ds^2=-A(r)dt^2+B(r)^{-1}dr^2+r^2d\theta^2+r^2\sin\theta\, d\varphi^2\,,
    \label{eq:metric}
\end{equation}
where the metric functions $A\,,B$ depend only on the radial coordiate $r$.

Our goal in this work is to examine the effect of the various terms appearing in the general shift-symmetric theory \eqref{eq:Horndeski} on the black hole QNM spectrum.
In general this is not a trivial task as one may consider arbitrarily higher order shift-symmetric terms by appropriately choosing the functions $G_i$.
As emphasized earlier, beyond GR solutions are found provided that a non-trivial coupling between the scalar field and the GB invariant is considered. This is ensured by the proper choice of the function $G_5$. For the remaining functions $G_i$ we consider leading order contributions in addition to the Einstein-Hilbert Lagrangian.
We consider the following theory \cite{Thaalba:2022bnt}
\begin{equation}
\begin{split}
    S=\frac{1}{2k}\int \mathrm{d}^4& x\,\sqrt{-g}\bigg[\frac{R}{2}+X+\alpha\,\phi\,\mathcal{G}\\
    &+\gamma \,G_{\mu\nu}\nabla^\mu\phi\nabla^\nu\phi
    +\sigma X\Box\phi +\kappa\, X^2 \bigg].
\end{split}
\label{eq:action}
\end{equation}
which may be obtained from action \eqref{eq:Horndeski} by selecting
\begin{equation}
\begin{split}
    &
    G_2(X) \coloneqq X+\kappa X^2\quad , \quad
    G_3(X) \coloneqq -\sigma X\, ,\\
    &
    G_4(X) \coloneqq 1/2 + \gamma X\quad , \quad
    G_5(X) \coloneqq -4\alpha\ln|X|\, .
\end{split}
\label{eq:Gi_functions}
\end{equation}
At first glance it may not be clear how these choices for the functions $G_i$ result in  \eqref{eq:action} based on the definition of the Horndeski Lagrangian \eqref{eq:Horndeski}.
However, it has been shown that the Gauss-Bonnet term appears in the Lagrangian under this specific choice of the function $G_5$ \cite{Kobayashi:2011nu,Sotiriou:2014pfa}. Moreover, it is straightforward to see that $\int \mathrm{d}^4x \sqrt{-g}\,(\Box\phi)^2=\int \mathrm{d}^4x \sqrt{-g}\,\big[(\nabla_\mu\nabla_\nu\phi)^2+R_{\mu\nu}\nabla^\mu\phi\nabla^\nu\phi\big]+\int \text{total derivative}$, which justifies the choice of $G_4$.

Henceforth, whenever a tilde is used above a symbol it will denote normalization with respect to the black hole horizon radius $r_h$, \textit{e.g.} $\tilde{\alpha}=\alpha/r_h^2,\,\tilde{\sigma}=\sigma/r_h^2,\,\tilde{\kappa}=\kappa/r_h^2$. Moreover, in a later section we will be using the letter $\zeta$ to denote the dimensionless parameter $\alpha/M^2$.

\subsection{Background solutions}

\begin{figure*}[!t]
    \includegraphics[width=0.325\linewidth]{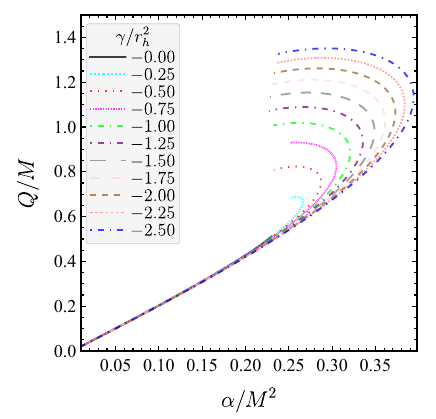}
    \includegraphics[width=0.325\linewidth]{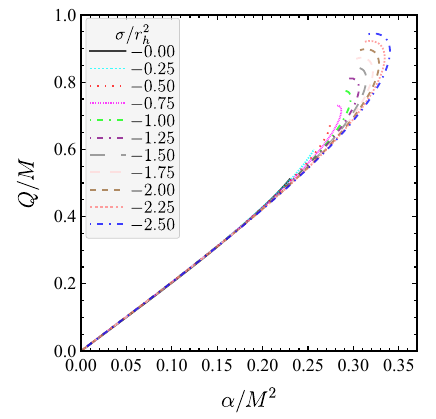}
    \includegraphics[width=0.325\linewidth]{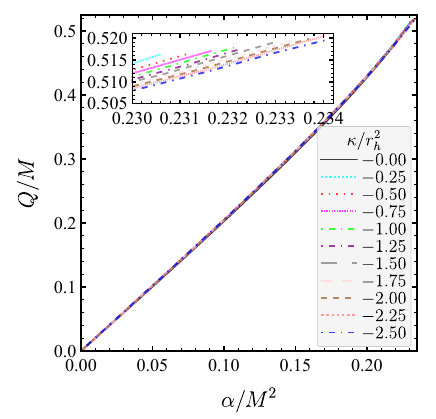}
    \caption{
    The scalar charge of the background solutions normalized with the black hole ADM mass for various $\mathcal{O}(1)$ negative couplings $\tilde{\gamma},\,\tilde{\sigma},\,\tilde{\kappa}$.
    The left plot corresponds to the case $\sigma=\kappa=0$, the middle one to $\gamma=\kappa=0$, and the right one to $\gamma=\sigma=0$.
    We see that the $\gamma$-term contribution has the most dominant effect, leading to larger scalar charges, and a larger parameter space of solutions which in turn allows for same mass black holes with different charges. The $\kappa$-term contribution has a very small effect on the properties of the hairy solutions as is demonstrated by the inset in the third plot.
    }
    \label{fig:background}
\end{figure*}

The Lagrangian we introduced in \eqref{eq:action} has been studied in \cite{Thaalba:2022bnt} under the choice for the functions given in \eqref{eq:Gi_functions}, where the nonminimal effects of the additional shift symmetric couplings on the properties of the solutions were studied. As has been pointed out there, regular black hole solutions may be found provided that the following existence conditions hold simultaneously at the black hole horizon:
\begin{align}
    \label{eq:condition_1}
    \textbf{I:}\quad & r_h^6-576 \alpha ^2 \gamma -24 \alpha  r_h^2 (8 \alpha +\sigma ) \ge 0\,,\\
    \label{eq:condition_2}
    \begin{split}
        \textbf{II:}\quad & \sqrt{r_h^6-576 \alpha ^2 \gamma -24 \alpha  (8 \alpha +\sigma ) r_h^2}\\
        &\times\big[24 \alpha  \gamma  r_h+(4 \alpha +\sigma ) r_h^3\big]
        +4\alpha\big[r_h^6-576 \alpha ^2 \gamma\\
        &\qquad-24 \alpha  (8 \alpha +\sigma ) r_h^2 \big]\ne 0\, .
    \end{split}
\end{align}
The combination of the conditions above reduces to the one presented in \cite{Sotiriou:2014pfa} in the limiting case $\{\gamma,\sigma\}\to \{0,0\}$.

Provided that the existence conditions are satisfied, numerical black hole solutions can be found presenting a black hole horizon at $r=r_h$ and being asymptotically flat with the spacetime approaching the Minkowski one when $r\to\infty$. 
Near the event horizon the metric elements assume the following expansions:
\begin{align}
    & A(r)=\sum_{n=1}a^{(n)}(r-r_h)^n\; ,\\
    & B(r)=\sum_{n=1}b^{(n)}(r-r_h)^n \; ,\\
    & \phi_0(r)=\sum_{n=0}\phi^{(n)}(r-r_h)^n\, ,
\end{align}
where the coefficients $\{a^{(n)},b^{(n)},c^{(n)}\}$ are obtained by inserting the above expansions into the field equations~\eqref{eq:bg_grav}-\eqref{eq:bg_scalar}, expanding close to the horizon radius, and solving order by order iteratively.
The couplings $\alpha$ and $\gamma$ contribute already at order $n=1$ to the boundary conditions, but $\sigma$ and $\kappa$ contribute only to higher orders.
Far from the BH, \textit{i.e.} for $r\gg r_h$, we have
\begin{align}
    & A(r)=1-\frac{2M}{r}+\sum_{n=2}\frac{A^{(n)}}{r^n} \; , \label{eq:expBgInfA}\\
    & B(r)=1-\frac{2M}{r}+\sum_{n=2}\frac{B^{(n)}}{r^n} \; , \label{eq:expBgInfB}\\
    & \phi_0(r)=\frac{Q}{r}+\sum_{n=2}\frac{\Phi^{(n)}}{r^n}\, , \label{eq:expBgInfphi}
\end{align}
where $M$ and $Q$ are the ADM mass and scalar charge  respectively. The above equations are consistent with the requirement that the BH solutions are asymptotically flat. Once again, the coefficients $\{A^{(n)},B^{(n)},\Phi^{(n)}\}$ can be obtained by substituting $A(r), B(r)$ and $\phi(r)$ into  Eqs.~\eqref{eq:bg_grav}-\eqref{eq:bg_scalar} and solving  order by order in $1/r$.

We numerically integrate the background equations and derive hairy asymptotically flat solutions. In order to calculate the ADM mass $M$ and scalar charge $Q$ we find the $1/r$ coefficients in the asymptotic expansions of the $g_{rr}$ metric component and scalar field respectively.
In this work we take $\gamma,\sigma,\kappa<0$ since these sign choices were shown in \cite{Thaalba:2022bnt} to have the most important effects on the solutions.
This can be indirectly inferred from the existence conditions \eqref{eq:condition_1}-\eqref{eq:condition_2}. For positive couplings $\gamma,\,\sigma$ the parameter space of solutions is generally suppressed. Since larger deviations (larger scalar charges) are found toward the edges of the allowed parameter space, \textit{i.e.} when $\alpha/M^2$ is maximized, the cases $\gamma,\,\sigma,>0$ are not expected to yield significant deviations from GR. The term involving $\kappa$ was shown in \cite{Thaalba:2022bnt} to follow a similar trend, albeit not directly entering the existence conditions.

In order to simplify our discussion, from this point onward when using the term ``existence line'' we will be referring to the continuous line running along the parameter space $[0,\alpha_\text{max}/M^2]$ which allows for hairy black hole solutions under specific choices of the parameters $\gamma,\,\sigma$ and $\kappa$.

The results we find confirm the those of \cite{Thaalba:2022bnt} and extend them.
We emphasize the following points which can be visually confirmed by inspecting Fig.~\ref{fig:background}:
first, despite not being directly related with the sourcing of the scalar hair, the additional shift-symmetric $\gamma,\,\sigma$ terms, affect significantly the minimum black hole mass and the amount of scalar charge the solutions carry.    
Second, in the small mass region of the parameter space (\textit{i.e.} large $\alpha/M^2$) for each $\gamma,\,\sigma$, one may find same-mass black holes that carry different scalar charges.
A way to interpret the meaning of the phrase ``same-mass black holes'' is to first consider a fixed value for the coupling $\alpha$, in which case moving along the horizontal axis of the plots in Fig.~\ref{fig:background} is equivalent to picking a different value for the black hole mass. Then black holes of a certain mass are found by considering the intersection of the existence line with a vertical line corresponding to that particular mass.
The solutions we present are derived while keeping the parameters $\tilde{\gamma}$ and $\tilde{\sigma}$ close to $\mathcal{O}(1)$, in accordance with the analysis done in \cite{Thaalba:2022bnt}.
It is clear that the effect that the $\gamma$ term has on the solutions is stronger in comparison with the $\sigma$ one, both on the minimum black hole mass and on the maximum scalar charge.

As mentioned earlier, $\kappa$ does not enter the existence conditions which are derived at the horizon. However, positive values of $\kappa$ may be responsible for the appearance of divergences at some distance $r>r_h$, which reduce the parameter space of solutions.
In any case,  $\kappa$ does not significantly alter the characteristics of the black holes in the region of the parameter space where solutions can be found, and therefore in the context of the QNMs we do not expect it to have a large impact.
We see the very small deviations that the $\kappa$ term introduces in the third panel of Fig.~\ref{fig:background}. In the inset plot we show a zoomed version of the tip of the curves so as to clarify the fact that $\kappa$ results in very small changes of the black hole mass and scalar charge.

\section{Perturbations}
\label{sec:perturbations}

\begin{figure*}[!t]
    \includegraphics[width=0.45\linewidth]{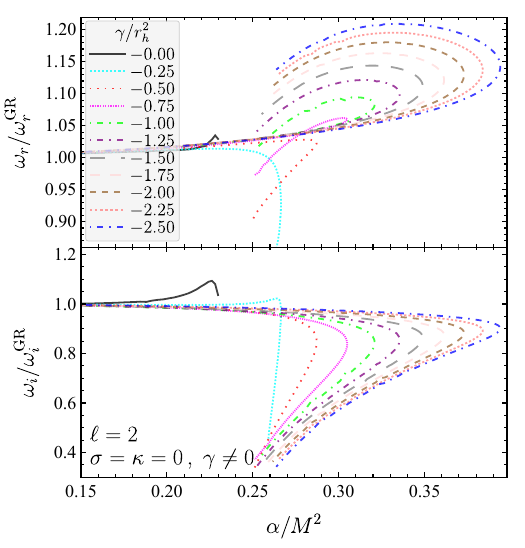}
    \includegraphics[width=0.45\linewidth]{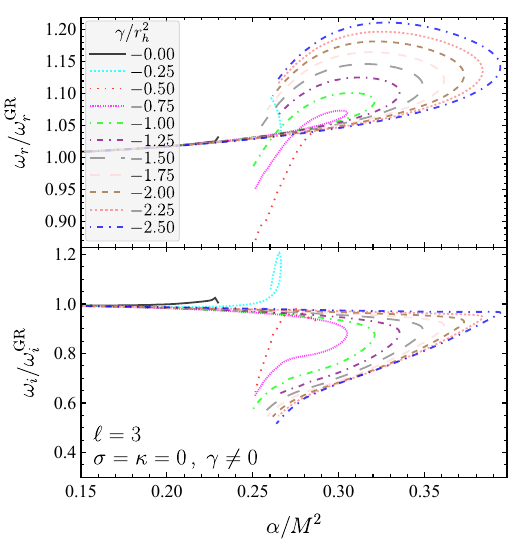}
    \caption{
    \textit{Left:}
    Real and imaginary parts of the axial QNM frequencies for angular number $\ell=2$ when $\alpha\ne 0$, $\gamma<0$ and $\sigma=\kappa=0$. The maximum value for $\omega_r$ increases while increasing $\gamma$ while the range of $\omega_i$, even though significantly different from the case $\gamma=0$, remains relative constant for $\gamma< 0$. The frequencies are normalized with respect to their GR counterparts.
    \textit{Right:}
    Same as the left plot but for $\ell=3$.
    }
    \label{fig:axial_alpha_gamma_negative}
\end{figure*}

We now introduce linear perturbations around the background solutions presented in the previous section. The metric and scalar perturbations at the linear level are given by:
\begin{align}
g_{\mu\nu}&=g_{\mu\nu}^{(0)}+\varepsilon\, \delta h_{\mu\nu}\,,\\
\phi&=\phi^{(0)}+\varepsilon\,\delta\phi\,,
\end{align}
where $\varepsilon \ll 1$ is a bookkeeping parameter. The quantities with superscript $(0)$ correspond to the numerical spherically symmetric background solutions which depend only on the radial coordinate $r$, while the perturbations are not spherically symmetric but rather functions of the spacetime coordinates $(t,r,\theta,\varphi)$.
As is usually done, we decompose the perturbations into the basis of spherical harmonics $Y_\ell^m(\theta,\varphi)$ which allows us to decouple them into the axial and polar sectors based on their properties under parity transformations \cite{Regge:1957td,Zerilli:1970se}. For polar perturbations the spherical harmonics transform according to $\hat{\textbf{P}}Y_\ell^m(\theta,\varphi)\to (-1)^\ell Y_\ell^m(\theta,\varphi)$, while for axial ones we have $\hat{\textbf{P}}Y_\ell^m(\theta,\varphi)\to (-1)^{\ell+1} Y_\ell^m(\theta,\varphi)$.

For the most part our analysis will be performed in the frequency domain where we will be employing the following decomposition of the perturbations:
\begin{align}
    \delta h_{\mu\nu}(t,r,\theta,\varphi)=& \int d\omega\, h_{\mu\nu}(r) Y_{\ell}^{m}(\theta,\varphi) e^{-i\omega t}\,,\label{eq:metric_decomposition}
    \\[2mm] 
    \delta\phi(t,r,\theta,\varphi)=& \int d\omega\, \frac{\phi_1(r)}{r}Y_{\ell}^{m}(\theta,\varphi) e^{-i\omega t}\, .\label{eq:scalar_decomposition}
\end{align}

Based on the preceding discussion in Sec.~\ref{sec:background}, we will only search for the effects of negative couplings $\gamma,\sigma,\kappa$ on the QNMs of the hairy black holes, as these are expected to have the most interesting effects on the QNM spectrum.
Since the frequency is complex, \textit{i.e.} $\omega=\omega_r+\omega_i\, i$ it is important that we check whether or not the sign of the imaginary part $\omega_i$ remains the same as in GR throughout the allowed parameter space.

\subsection{Axial modes}

If we adopt the Reggee-Wheeler gauge, we may write the axial perturbation components as follows:
\begin{equation}
    h_{\mu\nu}^a=
    \begin{bmatrix}
    0 & 0 & 0 & \sin\theta\,h_0\\
    0 & 0 & 0 & \sin\theta\, h_1\\
    0 & 0 & 0 & 0\\
    \sin\theta\,h_0 & \sin\theta\,h_1 & 0 & 0
    \end{bmatrix}
    \partial_\theta Y^{lm} \, ,
\end{equation}
where $(h_1,h_0)$ are functions of the radial coordinate $r$.
By eliminating one of the two functions appearing in the axial sector, we may arrive at the following partial differential equation which involves second derivatives with respect to time and radial coordinates:
\begin{equation}
   g(r)^2\frac{\partial^2 h_1}{\partial t^2} -\frac{\partial^2 h_1}{\partial r^2}+C(r)\frac{\partial h_1}{\partial r}+U(r)h_1=0 \, .
   \label{eq:wave_h1}
\end{equation}

For the theory described by \eqref{eq:action}, we find that apart from the linear GB interaction, only the $\gamma$ coupling enters the function $g(r)$ directly. We may use the function $g(r)$ to define the tortoise coordinate $r_*$:
\begin{equation}
    \frac{dr_*}{dr}\equiv g(r)=\sqrt{\frac{1-4 \alpha  B' \phi'-B (8 \alpha  \phi''+\gamma  \phi'^2)}{B (-4 \alpha  B A' \phi'+A B \gamma  \phi'^2+A)}}\, .
\end{equation}
However, it is important to stress that the background solutions for $A(r),\,B(r)$ and $\phi(r)$ are not only affected by $\alpha$ and $\gamma$, but also by $\sigma$ and $\kappa$, which means that all the additional shift-symmetric terms affect the behaviour of the function $g(r)$ indirectly.
By inspecting the form of \eqref{eq:wave_h1} we deduce that the sign of the function $g(r)$ determines the hyperbolic nature of the equation. In what follows we will be examining the hyperbolicity of the equation and the point at which it is lost. This will serve as the upper limit for our calculations of the QNMs. 

\begin{figure*}[!t]
    \centering
    \includegraphics[width=0.45\linewidth]{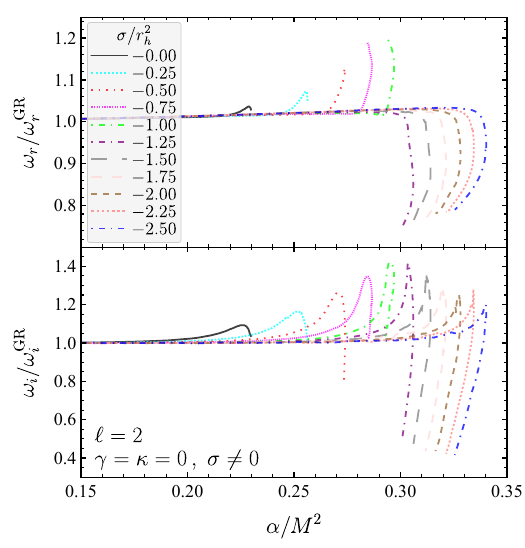}
    \includegraphics[width=0.45\linewidth]{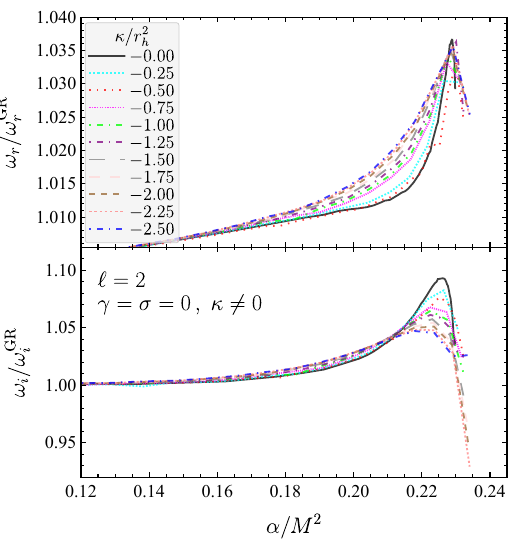}
    \caption{\textit{Left:}
    Real and imaginary parts of the axial QNM frequencies for angular number $\ell=2$ when $\alpha\ne 0$, $\sigma<0$ and $\gamma=\kappa=0$. The most prominent deviations occur toward the edge of the existence lines which corresponds to the small mass limit.
    \textit{Right:} Same as the left plot but for $\alpha\ne 0$, $\kappa<0$ and $\gamma=\sigma=0$. In this case, similarly to what happens with the characteristics of the stationary solutions, the results are only slightly affected from the value of $\tilde{\kappa}$.
    }
    \label{fig:axial_alpha_sigma_kappa_negative}
\end{figure*}

After performing the decomposition \eqref{eq:metric_decomposition}-\eqref{eq:scalar_decomposition}, we may define the following vectors in terms of the functions appearing in the perturbations, along with the appropriate boundary conditions for the QNMs:
\begin{equation}
    \boldsymbol{\Psi}^a\equiv (h_0,h_1)^\top \quad , \quad \lim_{r_*\to r_*^{\pm}}\boldsymbol{\Psi}^{a} \propto e^{\pm i\omega r_*}
\; , \;  
\end{equation}
The boundary conditions for QNMs correspond to purely outgoing waves at infinity and purely ingoing waves at the horizon, where $r_*$ is the ``tortoise'' coordinate.
The equations for $h_1$ and $h_0$ can then be written in a compact matrix form as:
\begin{equation}
    \frac{d}{dr}\boldsymbol{\Psi}^a_i+C^a_{ij}\,\boldsymbol{\Psi}^a_j=0\, ,
\label{eq:wave_Psi}
\end{equation}
where $\boldsymbol{\Psi}^a= (h_0,h_1)^\top$, and the coefficients $C^a_{ij}$ depend on $\omega$ and on the numerical background.
We solve Eqs.~\eqref{eq:wave_Psi} trough the so called matrix method. 
We first impose suitable boundary conditions:
\begin{align}
    {\bf \Psi}_i^a(r_\star\rightarrow -\infty)=&\sum_{n_\textnormal{min}}^{n_\textnormal{max}} c_{i}^{(n)}e^{-i\omega r_\star}(r-r_h)^n\,,\label{eq:BChor}\\
    {\bf \Psi}_i^a(r_\star\rightarrow+ \infty)=&\sum_{n_\textnormal{min}}^{n_\textnormal{max}} \frac{\bar{c}_{i}^{(n)}}{r^n}e^{i\omega r_\star}\, .\label{eq:BCinf}
\end{align}
For the numerical integration we properly fix the expansion order at the horizon $n_\text{min}$ and at asymptotic infinity $n_\text{max}$, making sure that our numerical results do not improve significantly for $n>n_\text{max}$. 
We then create the $2\times2$ matrix given by two independent solutions of Eq.~\eqref{eq:wave_Psi}:
\begin{equation}
 {\textbf X}_a= \big(\boldsymbol{\Psi}^a_{(h)},{\bf \Psi}^a_{(\infty)}\big)
= \begin{bmatrix}
    h_0^{(h)} & h_0^{(\infty)}\\
    h_1^{(h)} & h_1^{(\infty)}\\
 \end{bmatrix}\ .
\end{equation}
The first solution denoted with an $(h)$ superscript is obtained by integrating the differential equations \eqref{eq:wave_Psi} from the horizon outward with boundary condition \eqref{eq:BChor}, 
while the second one denoted with $(\infty)$ is determined by integrating from infinity inward using as boundary conditions Eqs.~\eqref{eq:BCinf}. QNM frequencies correspond to poles of the determinant of the fundamental matrix, \textit{i.e.} they are given by solving:
\begin{equation}
\textnormal{det}\ \textbf{X}(\omega)|_{r_m}=0\, ,
\end{equation}
While performing the shooting method we are checking that our results are independent of the choice for the intermediate matching point radius $r_m$.

\subsubsection{Frequency domain analysis}

We want to examine the effect of each one of the three additional shift-symmetric terms, and therefore we will consider their contributions to the QNMs separately.
For the most part we examine the results for angular number $\ell=2$, as we are interested in the dominant modes. In GR the mode for $\ell=2$ corresponds to $M \omega^{\text{GR}}\approx 0.3737 - 0.08896\, i$, while for $\ell=3$ the value of the mode is $M \omega^{\text{GR}}\approx 0.5994 - 0.09270$.

First, in Fig.~\ref{fig:axial_alpha_gamma_negative} we show the axial mode QNM frequencies normalized with respect to the GR modes, in the scenario where $\gamma<0$ and $\sigma=\kappa=0$.
The solid line in the figures corresponds to the QNMs when only the GB coupling is considered, and it clearly leads to very small differences with respect to the GR case. Specifically, for $\ell=2$, the difference in the real part of the frequency is $\sim 3 \%$ at most, towards the edge of the existence line. The imaginary part can differ by as much as $\sim 10 \%$ in the respective limit of the curve.
Things change significantly however, when $\gamma<0$ is introduced. In Fig.~\ref{fig:axial_alpha_gamma_negative} we take $\tilde{\gamma}$ to be as negative as $-2.5$, in which case $\omega_r$ becomes as large as $\sim 1.21\times \omega_r^\text{GR}$, a result which is considerably larger in comparison with the mere $3\%$ increase in its absence.
One more thing of particular importance relates to the fact that now, in order to observe significant deviations from GR, one needs not to look toward the edge of the existence curve.
In fact, the maximum deviations of the oscillation frequencies are observed around the middle of the allowed parameter space. Let us also note that the overall effect that the negative $\gamma$ term has on $\omega_r$ is to increase it, while on $\omega_i$ is the opposite.
The results for $\ell=3$ are similar, with the only obvious difference being that $\omega_i$ changes more moderately as one moves along the existence line.

We then proceed with the same analysis examining now the $\sigma$ contribution instead of the $\gamma$ one. The results are presented in the left panel of Fig.~\ref{fig:axial_alpha_sigma_kappa_negative} for $\ell=2$.
The main difference we observe with regards to the $\gamma$ contribution, is that significant deviations from GR are only exhibited very close to the edge of the existence line, \textit{i.e.} in the small mass limit of each branch.
In particular, we notice that for $\tilde{\sigma}\ge-1$ the oscillation frequency becomes larger than that of GR in the small mass limit, while the opposite occurs when $\tilde{\sigma}\le -1.25$.
As for the imaginary part, we notice a trend for all choices of $\tilde{\sigma}$, according to which for larger and intermediate masses, $\omega_i/\omega_i^\text{GR}$ increases, and it starts decreasing in the small mass limit towards the edge of the existence line.
Like in the $\gamma$ case, we deduce that in contrast with the sole $\alpha$ coupling scenario, the $\sigma$ contribution results in significant deviations from GR (even though, as already mentioned, the large effects are now limited to small masses).

\begin{figure}[!t]
    \includegraphics[width=0.95\linewidth]{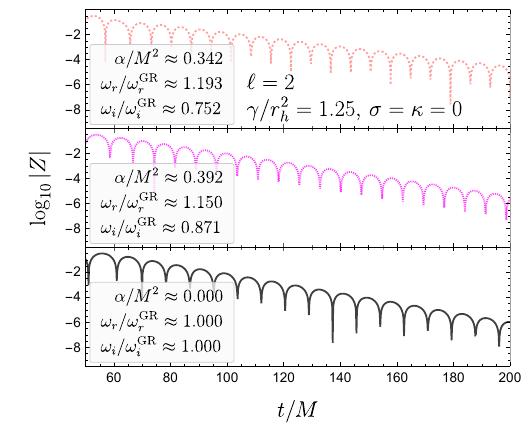}
    \caption{
    Time domain profile of the axial gravitational perturbation signal $Z(v,u)$ in the scenario with $\tilde{\gamma}=2.50,\,\sigma=\kappa=0$ and $\ell=2$ for $\alpha/M^2\approx 0.342,\, 0.392,\, 0$.
    As initial condition we assume a static Gaussian pulse with parameters $A=1,\,w=5 M$ and $v_0=25 M$, where $M$ is the hairy black hole's mass. The real and imaginary parts of the mode are retrieved by fitting the signal profiles with damped sinusoids.
    }
    \label{fig:time_evolution}
\end{figure}

From Fig.~\ref{fig:background} which demonstrates that the $\kappa$ term has a small effect on the background solutions, we expect that it will not to be as influential on the QNMs as the $\gamma$ and $\sigma$ terms.
Indeed, the right panel of Fig.~\ref{fig:axial_alpha_sigma_kappa_negative} shows that $\kappa$ leads to only slightly faster oscillating solutions for the most part of the parameter space. The decay rate is also almost unaffected for the most part and only slightly smaller in the small mass limit.

A common trait between all different scenarios is that despite the imaginary part of the frequency decreasing for most couplings, especially toward the ends of the existence lines, it never reaches zero.
Overall the axial mode results suggest that linear stability is a generic characteristic of the solutions.

\subsubsection{Time domain analysis}

For the axial sector it is relatively simple to work in the time domain instead of the frequency one, and solve directly the partial differential equation \eqref{eq:wave_h1}.
To that extend, it is easier to work in a set of lightcone 
coordinates \cite{Gundlach:1993tp,Konoplya:2011qq} 
$u=t-r_*$ and $v=t+r_*$.
The first step is to re-write the perturbation equation \eqref{eq:wave_h1} in terms of the tortoise coordinate\footnote{Applying the frequency decomposition to this equation results in a Schr\"odinger-type equation with the potential being the one of axial perturbations.}:
\begin{equation}
    -\frac{\partial^2 h_1}{\partial t^2}+\frac{\partial^2 h_1}{\partial r_*^2}+V_a(t,r)h_1=0 \, ,
\end{equation}
which after substituting the lightcone coordinates becomes:
\begin{equation}
\bigg[4\,\frac{\partial^2}{\partial u\, \partial v}+V_a(u,v)\bigg]Z(u,v)=0\, .
\end{equation}
Solving the above equation with appropriate boundary conditions allows us to extract the QNM frequency by appropriately fitting the resulting signal. This process is obviously less effective than the frequency domain integration if we want to span the whole parameter space of solutions, but may serve as a confirmation of the results extracted from the analysis in the previous subsection.

Here we will examine an example by considering as initial condition a static Gaussian pulse, \textit{i.e.},:
\begin{equation}
    Z(0,v)=A\exp\bigg[-\frac{(v-v_0)^2}{2w^2}\bigg]\, , \; Z(u,0)=0\, .
\end{equation}
For the numerical calculations the width of the Gaussian pulse is set to be $w=5M$ while $v_0=25M$, where $M$ is the mass of the hairy black hole. Of course the results should be independent of these choices.
For convenience we also set the amplitude to $A=1$.
The results of the example are shown in Fig.~\ref{fig:time_evolution}, where we present the signal obtained at a distance of $r=25M$ for three separate black hole solutions (\textit{i.e.} different values of $\alpha/M^2$) in the scenario with $\tilde{\gamma}=2.50,\,\sigma=\kappa=0$ and $\ell=2$. The results for the QNMs obtained by appropriately fitting the signal are consistent with the results acquired by the frequency domain integration which were presented in Fig.~\ref{fig:axial_alpha_gamma_negative}.
Notice that the signal in the top panel for $\alpha/M^2\approx 0.342$ corresponds to the black hole solution with the larger scalar charge (as we explained earlier going beyond a certain value of $\alpha/M^2$ for each $\tilde{\gamma}$ allows for two differently charged solutions).

\subsection{Polar modes}

\begin{figure*}[!t]
    {\Large \textbf{Polar gravitational-led modes for $\boldsymbol{\ell=2}$}}\\[2mm]
    \includegraphics[width=0.95\linewidth]{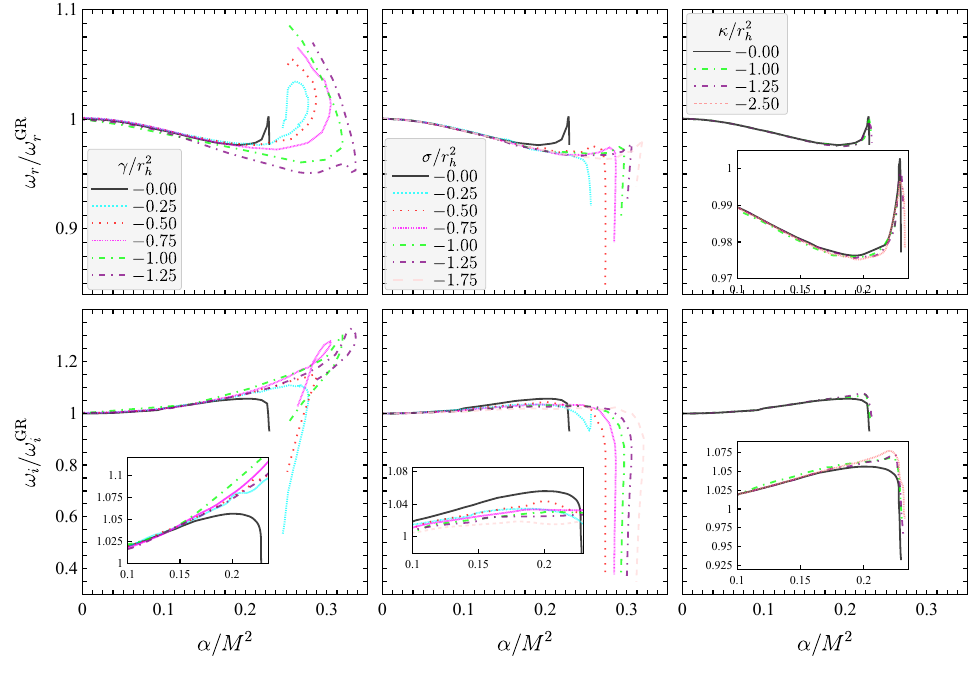}
    \caption{
    Here we present the normalized real and imaginary parts of the QNM frequencies for the polar gravitational-led modes. The left column corresponds to $\gamma\ne 0$, the middle one to $\sigma\ne 0$, and the right column to $\kappa\ne 0$. The vertical and horizontal scales have been kept consistent between the different scenarios in order to allow for a direct visual comparison. For better legibility we also include some zoomed plots.
    }
    \label{fig:polar_grav}
\end{figure*}

\begin{figure*}[!t]
    \centering
    {\Large \textbf{Polar scalar-led modes for $\boldsymbol{\ell=2}$}}\\[2mm]
    \includegraphics[width=0.95\linewidth]{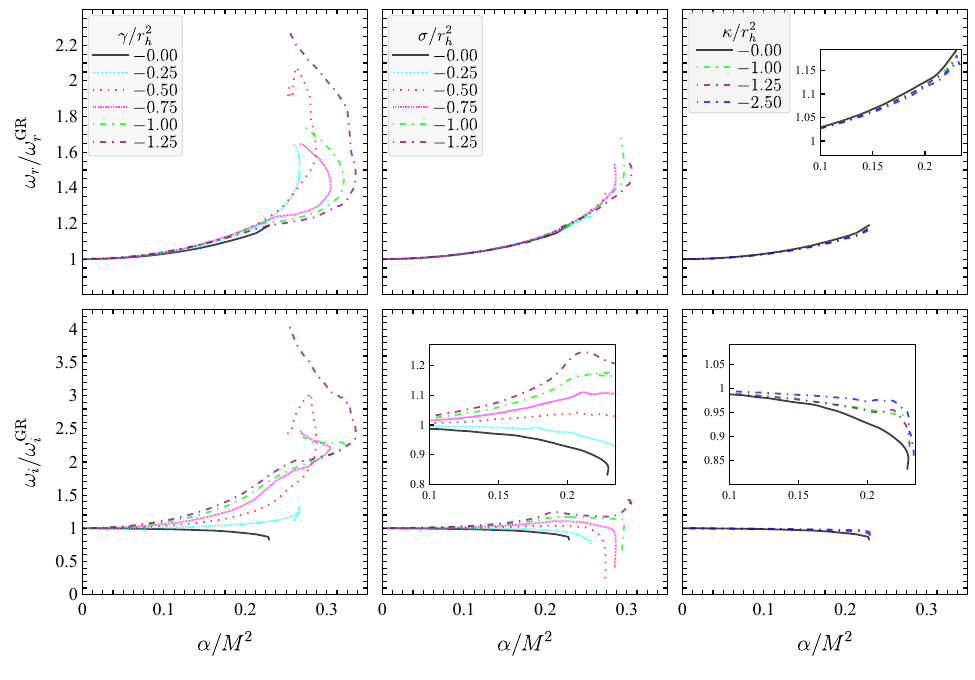}
    \caption{
    The normalized real and imaginary parts of the QNM frequencies for the polar scalar-led modes. The left column corresponds to $\gamma\ne 0$, the middle one to $\sigma\ne 0$, and the right column to $\kappa\ne 0$, where once again the vertical and horizontal scales have been kept consistent between the different scenarios in order to allow for a direct visual comparison. We also include some zoomed plots to make some of the results easier to read.
    }
    \label{fig:polar_scalar}
\end{figure*}

We may now move on to study the polar perturbations.
The metric perturbations in the polar sector may be expressed as:
\begin{equation}
    h_{\mu\nu}^p=
    \begin{bmatrix}
    A H_0 &  H_1  & 0 & 0 \\
    H_1 & H_2/B & 0 & 0\\
    0 & 0 & r^2 K & 0\\
    0 & 0 & 0 & r^2 \sin^2 \theta \, K
    \end{bmatrix}
    Y^{lm}\, ,
\end{equation}
where $(H_2,H_1,H_0,K,)$ are spherically symmetric functions.
In analogy with the axial modes analysis, for the study of the polar QNMs we take:
\begin{equation}
    \boldsymbol{\Psi}^p\equiv  (K,H_1,\phi_1,\phi_1')^\top \quad , \quad \lim_{r_*\to r_*^{\pm}}\boldsymbol{\Psi}^{p} \propto e^{\pm i\omega r_*}
\; , \;  
\end{equation}

In the polar scenario we may in principle recover two types of QNMs, namely the gravitational-led modes and the scalar-led ones, since now the scalar field perturbations do not decouple as in the axial scenario.
By properly handling the equations we may write the polar perturbation problem in a compact matrix form:
\begin{equation}
    \frac{d}{dr}\boldsymbol{\Psi}^p_i+C^p_{ij}\,\boldsymbol{\Psi}^p_j=0\, ,
\label{eq:polar_wave_Psi}
\end{equation}
where $\boldsymbol{\Psi}^p= (K,H_1,\phi_1,\phi_1')^\top$ and the coefficients $C^p_{ij}$ depend on $\omega$ and the numerical background.
Once again we consider appropriate expansions for the perturbations at the horizon and infinity:
\begin{align}
    {\bf \Psi}_i^p(r_\star\rightarrow -\infty)=&\sum_{n_\textnormal{min}}^{n_\textnormal{max}} d_{i}^{(n)}e^{-i\omega r_\star}(r-r_h)^n\, ,\\
    {\bf \Psi}_i^p(r_\star\rightarrow+ \infty)=&\sum_{n_\textnormal{min}}^{n_\textnormal{max}} \frac{\bar{d}_{i}^{(n)}}{r^n}e^{i\omega r_\star}\, .
\end{align}
Once again we ensure that we have considered a sufficiently high order, so that the results of the integration remain unchanged if we further increase the order of the expansions.

We now create the following $4\times 4$ matrix:
\begin{equation}
\begin{split}
     {\textbf X}_p= & \;\big(\boldsymbol{\Psi}^p_{(h,1)},\boldsymbol{\Psi}^p_{(h,2)},\boldsymbol{\Psi}^p_{(\infty,1)},\boldsymbol{\Psi}^p_{(\infty,2)}\big)\\[2mm]
        = &
    \begin{bmatrix}
    K^{(h,1)} & K^{(h,2)} & K^{(\infty,1)} & K^{(\infty,2)}\\
    H_1^{(h,1)} & H_1^{(h,1)} & H_1^{(\infty,1)} & H_1^{(\infty,2)}\\
    \phi_1^{(h,1)} & \phi_1^{(h,1)} & \phi_1^{(\infty,1)} & \phi_1^{(\infty,2)}\\
    \phi_1'^{(h,1)} & \phi_1'^{(h,1)} & \phi_1'^{(\infty,1)} & \phi_1'^{(\infty,2)}\\
    \end{bmatrix}\, ,
\end{split}
\end{equation}
which is made of independent solutions of \eqref{eq:polar_wave_Psi}, derived by integrations from the horizon outward and from infinity inward.
Solving the system of differential equations is now far more computationally demanding than in the axial case. This relates both to the larger number of differential equations (four instead of two), and also to the far more complicated expressions characterizing the boundary conditions, which now correspond to three functions ($H_1,\,K,\,\phi_1$) instead of two ($h_0,\,h_1$).
Perhaps most importantly for computations, increased accuracy is required particularly in the small mass regime where the scalar field contribution becomes especially important, as opposed to the axial case where the scalar and gravitational perturbations decouple.

Similarly to the axial case, we study the polar QNMs by focusing on the contributions of the $\gamma,\,\sigma$ and $\kappa$ terms separately. We group the results in Fig.~\ref{fig:polar_grav} and Fig.~\ref{fig:polar_scalar}, one for the gravitational-led modes and one for the scalar-led ones. Once again we focus on angular number $\ell=2$.
The first and second rows of plots in each figure depict the normalized real and imaginary parts of the frequency respectively. Each one of the three columns in the figures examines the effects of $\gamma,\,\sigma$ and $\kappa$ separately. Let us also note that we chose the same vertical and horizontal range for each one of the three different scenarios so as to allow for direct visual comparison of the effects the different terms cause.
For the plots in these subsection, we normalize with respect to the GR value for gravitational perturbations with angular number $\ell=2$ which was given in the subsection about the axial modes, and for scalar perturbations with $\ell=2$, which is $M \omega^{\text{GR}}\approx 0.4836 - 0.09676$.

Let us first consider the effects of the additional terms on the gravitational-led polar perturbations.
From the first column of Fig.~\ref{fig:polar_grav} we observe that unlike in the simpler model where only $\alpha\ne 0$, introducing a nonzero negative $\gamma$ leads to considerably large deviations from GR. In particular for intermediate values of $\alpha/M^2$ in the according allowed range, the frequency of oscillations decreases and the damping rate increases with respect to GR.
Moving into the higher $\alpha/M^2$ region of the parameter space, same-mass black holes with different charges exist. For the black holes with the smaller charge the aforementioned trend continuous, but for the black holes with the larger charges the behaviour can be reversed, \textit{i.e.} $\omega_r>\omega_r^\text{GR}$ and $\omega_i<\omega_i^\text{GR}$.
When $\sigma< 0$ it turns out that the numerical integration requires higher accuracy and smaller integration steps, making the computations much more time consuming than in the $\gamma< 0$ case.
Based on our earlier results concerning the background and axial mode solutions, we do not expect the $\sigma$ term to have more prominent effects on the modes compared with $\gamma$, especially in the intermediate mass range.
Indeed, from the second column of Fig.~\ref{fig:polar_grav}, we deduce that the effects of the $\sigma$ term are limited to the edge of the allowed parameter space and are far more prevalent in the decay rate which can be significantly reduced (by as much as $\sim 60\%$ for $\tilde{\sigma}\lesssim -0.5$). Regarding the contributions of the $\kappa$ interaction, we see that the effects are very small.

Let us now move on to the study of the scalar-led polar perturbations, while once again focusing on the case with angular number $\ell=2$.
The additional terms have a far more pronounced effect on the scalar-led modes as can be easily seen by inspecting Fig.~\ref{fig:polar_scalar}.
Out of the three different terms, the $\gamma$ one produces the most dominant contributions. For $\gamma\lesssim -0.5$ the frequency of oscillations may even double for large values of $\alpha/M^2$, while the imaginary part of the frequency may triple or even quadruple with respect to GR. Overall, negative values of $\gamma$ lead to an increase on both the real and imaginary part of the frequencies.
As for the $\sigma$ contribution, we discern that negative couplings lead to an increase -albeit smaller than the one in the case of the $\gamma$ term- of the oscillation frequency. The effects on the decay rate however, is not consistent for all couplings $\sigma$ and the behaviour of the ratio $\omega_i/\omega^\text{GR}_i$ is non-monotonic.
Finally, in the case of the $\kappa$-term, the effect on the modes is yet again limited, in alignment with the respective results for the axial and gravitational-led polar modes.

Like in the axial modes, the imaginary part of the frequency remains negative irrespective of the case we consider. This time however and in contrast to the axial scenario, there are cases when $\omega_i$ reduces significantly, especially in the end of the existence lines.
By comparing the decay rate of axial and polar modes we can see that overall, axial modes may be longer lived than the polar ones, which is in contrast to what happens in the simpler theory where $\gamma=\sigma=\kappa=0$.

\begin{table}
    \begin{tabular}{|c|c|}
    \hline\hline
    \multicolumn{2}{|c|}{\textbf{Axial modes $\boldsymbol{\ell=2,\; \alpha/M^2\approx 0.2}$}} \\
    \hline
    $(\tilde{\gamma},\tilde{\sigma},\tilde{\kappa})$
    & $\boldsymbol{M\omega}$ \\
    \hline\hline
    $(0,0,0)$ & $0.3779-0.09087\, i$ \\[2mm]
    $(-1,0,0)$ & $0.3795-0.08753\, i$ \\[2mm]
    $(0,-1,0)$ & $0.3786-0.08909\, i$ \\[2mm]
    $(0,0,-1)$ & $0.3785-0.09123\, i$ \\[2mm]
    $(-1,-1,-1)$ & $0.3791-0.08829\, i$ \\[2mm]
    \hline
\end{tabular}
    \caption{In this table we show the axial QNM frequencies for hairy black holes with $\alpha/M^2\approx 0.2$ for various $\mathcal{O}(1)$ combinations $(\tilde{\gamma},\tilde{\sigma},\tilde{\kappa})$. The most prominent effects of the additional shift-symmetric terms appear towards the end of the existence lines, and therefore despite these values being interesting in the context of comparing same-mass black holes existing, they do not yield very large deviations from GR.}
    \label{tab:table_axial_QNMs}
\end{table}

\begin{table}
    \begin{tabular}{|c||c|c|}
    \hline\hline
    \multicolumn{3}{|c|}{\textbf{Polar modes $\boldsymbol{\ell=2,\; \alpha/M^2\approx 0.2}$}} \\
    \hline
    $\boldsymbol{(\tilde{\gamma},\tilde{\sigma},\tilde{\kappa})}$
    & \textbf{grav-led} $\boldsymbol{M\omega}$ & \textbf{scalar-led} $\boldsymbol{M\omega}$  \\
    \hline\hline
    $(0,0,0)$ & $0.3652-0.09395\, i$ & $0.5437-0.08619\, i$ \\[2mm]
    $(-1,0,0)$ & $0.3520-0.1008\, i$ & $0.5626-0.1419\, i$ \\[2mm]
    $(0,-1,0)$ & $0.3628-0.09162\, i$ & $0.5494-0.1112\, i$ \\[2mm]
    $(0,0,-1)$ & $0.3647-0.09482\, i$ & $0.5453-0.09302\, i$ \\[2mm]
    $(-1,-1,-1)$ & $0.3619-0.09288\, i$ & $0.5443-0.1173\, i$ \\[2mm]
    \hline
\end{tabular}
    \caption{Same as Tab.~\ref{tab:table_axial_QNMs} but showing the polar grav-led and axial-led modes.}
    \label{tab:table_polar_QNMs}
\end{table}

Before closing this section, let us comment on differences on the QNMs arising in the various scenarios for same-mass black holes in different scenarios.
A reasonable question to ask is the following: among the three additional shift-symmetric terms we considered in this work, is there any one in particular that is expected to yield more important contributions than the rest?
As it became clear from the analysis of both the axial and polar spectrum, the $\kappa$-term has very limited effects on the QNMs in comparison with $\gamma$ and $\sigma$. This result was expected based on the effect that same term has on the scalar charge of the stationary solutions. Between $\gamma$ and $\sigma$ the effects were comparable towards the edge of the existence lines, but $\gamma$ seemed to exert a stronger effect on the QNMs for heavier black holes (\textit{i.e.} for smaller $\alpha/M^2$).
Since larger mass black holes are particularly interesting from an observation point of view (see Sec. \ref{sec:astrophysics}), the contribution of the $\gamma$ term is expected to be the most interesting in comparison with $\sigma$ and $\kappa$.
In Tab.~\ref{tab:table_axial_QNMs} and \ref{tab:table_polar_QNMs} we present the results for the QNMs in four different models characterized by the parameters $(\tilde{\gamma},\tilde{\sigma},\tilde{\kappa})$, with $\ell=2$ and $\alpha/M^2=0.2$.
The four different cases are supplemented with the simplest case where all the additional shift-symmetric couplings are ignored, which is shown in the first row of the tables. The fourth case we considered corresponds to an $\mathcal{O}(1)$ combination of all three terms.
Notice that in order to be able to compare same-mass black holes between the different models we have to resort to a value for $\alpha/M^2$ which yields hairy solutions for all models simultaneously. We have fixed that value to be $0.2$. From the table we see that $\gamma< 0$ appears to be the most influential among the four scenarios, even considering the mixed one with $(\tilde{\gamma},\tilde{\sigma},\tilde{\kappa})=(-1,-1,-1)$.
This further strengthens the conclusion that $\gamma$ is the most dominant of the terms we considered.

\section{Astrophysical implications}
\label{sec:astrophysics}

Finally, we should mention that there have been various works constraining the value of $\alpha$ in models such as dilatonic GB gravity or GR supplemented simply by the linear scalar-GB interaction \cite{Yagi:2012gp,Nair:2019iur,Yamada:2019zrb,Wang:2021jfc,Perkins:2021mhb,Saffer:2021gak,Lyu:2022gdr,Wong:2022wni}.
With our chosen notation these analyses yield constraints which for light black holes translate to $\mathcal{O}(\alpha/M^2)$ being one\footnote{In principle, one would need to see how the constraints are modified by the inclusion of the additional terms.}.
To be consistent with the majority of the bibliography we should convert our notation to the following:
\begin{equation}
\begin{split}
    S=&\int d^4x \sqrt{-g}\left[\frac{R}{2k}+\bar{X}+\bar{\alpha}\varphi\,\GB\right]=\\
    &
    \frac{1}{k}\int d^4x \sqrt{-g}\left[\frac{R}{2}+k\bar{X}+(\sqrt{k}\bar{\alpha})(\sqrt{k}\varphi)\,\GB\right]
    \,,
\end{split}
\end{equation}
so that the relation between the two notations is expressed through $\phi\equiv\sqrt{k}\,\varphi$ and $\alpha\equiv\sqrt{k}\,\bar{\alpha}$.
Then considering the stringest type of constraints for which $\sqrt{\bar{a}}\le\sqrt{\bar{a}_{\text{c}}}\approx 1\,\text{km}$, we see that in our notation the constraint value corresponds to $\sqrt{\alpha}_\text{c}\approx \sqrt{k} \,\text{km}$. Then, the constraint is satisfied for $\zeta<\alpha_\text{c}M^{-2}$.
In this context, $\zeta\equiv\alpha/M^2$ is the usual parameter we used in the horizontal axes of most of the plots throughout this work.
This condition yields a constraint value for $\zeta$ which after reintroducing the physical constants is given by
$\zeta_\text{c}\approx 2.3 \,(M_\odot/M)^2$.
Therefore, for light black holes of a few solar masses the deviations in the QNMs we observed due to the additional shift-symmetric terms can appear within the observational bounds.
If for example we take the lighter mass object of GW190814 \cite{LIGOScientific:2020zkf} with mass $2.6M_\odot$ to be a black hole, we would be looking at a value $\zeta\approx 0.34$, which in our analysis corresponds to a region of the parameter space where very large deviations from GR can occur when the additional shift-symmetric terms are considered. Of course this example only serves as an extreme one, but for $\sqrt{\alpha}\lesssim \sqrt{k}\,\text{km}$ even more massive black holes can lead to values of $\zeta$ which are consistent with relatively large deviations. As we can see from the various plots in the previous sections differences of the order of 1-10\% may even occur for $\zeta<0.1$ and couplings $\tilde{\gamma},\tilde{\sigma}\sim\mathcal{O}(1)$, with the largest deviations noticed for the scalar-led perturbations.

By comparing the GR QNM spectrum to the one presented in this work, one could potentially derive additional constraints on the value of $\alpha$ as well as on the other parameters of the theory, namely $\gamma,\,\sigma$ and $\kappa$. Deriving such constraints would be meaningful if one considers light black holes characterized by relatively large values of $\zeta$.
Recently, it was demonstrated that in the small scalar charge approximation (which in our analysis corresponds to a small $\alpha/M^2$ ratio), corrections to the QNMs are suppressed for very massive black holes relevant for LISA, since black holes at those mass scales are not expected to carry significant charges \cite{DAddario:2023erc}. This result is in agreement with the results derived in this work.
In either case, using the results for the QNM spectrum to place constraints on the parameters of the theory or test it against GR predictions is beyond the scope of the current work.

\section{Conclusions}
\label{sec:conclusions}

We have studied axial and polar perturbations of hairy black holes in shift-symmetric Horndeski gravity. The main goal of this work was to study how QNMs are impacted by shift-symmetric terms that appear in the theory as an addition to the hair-sourcing scalar-GB interaction. This was partly motivated by the study of such terms and their nontrivial impact on the stationary hairy background solutions \cite{Thaalba:2022bnt}.
To study the modes we divided our analysis into axial and polar perturbations.

For the axial modes we concluded that the $\gamma$ and $\sigma$ terms influence significantly both the oscillation frequency and the damping rate, while the effects of the $\kappa$ interaction are very limited. Despite working almost exclusively in the frequency domain, we examined an example of axial modes in the time domain and found that our results are in good agreement between the two methods.

For the polar modes the analysis presents extra complexity since now two types of modes can be found, namely gravitational-led and scalar-led ones. The isospectrality between axial and polar modes was broken. Similarly to the axial case we found the $\gamma$ and $\sigma$ terms to be the ones that predominantly affect the modes. Especially in the scalar-led modes the effects were very important both in the real and imaginary part of the frequency. One interesting characteristic that we observed relates to the fact that by including the $\sigma$ and $\gamma$ contributions, the imaginary part of the frequency was in general larger in the polar modes and not in the axial ones, which is in contrast to the case when we only consider the $\phi\GB$ interaction.

Overall, our analysis suggests that the $\gamma$ and $\sigma$ terms have the potential to significantly alter the QNM spectrum both in the axial and polar sector. In all scenarios the effects are larger the further away from GR one goes (\textit{i.e.} for larger $\alpha/M^2$).
Another point worth emphasizing relates to possible degeneracies arising in the different scenarios.
Based on the results of this work, it is reasonable to deduce that degeneracies may in principle exist between same-mass black holes which either belong in different models (\textit{e.g.} in $\gamma<0,\;\sigma=\kappa=0$ and $\sigma<0,\;\gamma=\sigma=0$), or are allowed within the same model. The latter is possible due to the existence of same-mass black holes carrying different scalar charges as we explained in Sec.~\ref{sec:background}.

Our results showed that the frequencies and decay rates of the modes may vary considerably depending on the exact model.
It would, therefore, be important to examine in detail the ability of future interferometers to access these effects. At the same time, a thorough analysis of potential new constraints on the parameters of the theory should be performed.
Finally, there is a need for an extensive analysis of the exact region of the parameter space where a perturbative approach in accordance with \cite{DAddario:2023erc} is possible. This will allow us to conclusively determine whether or not results such as the ones presented in the current work will be relevant for LISA.
These tasks go beyond the purpose of this paper and are left for future work.
Finally, a potential continuation of the current work could examine the grey-body factors in generic frameworks such as the shift-symmetric one studied here.

\acknowledgements
I am grateful to Thomas P. Sotiriou for reading the manuscript and making useful suggestions.
I acknowledge financial support by INFN through the post-doctoral fellowships 2023.

\appendix

\section{Background equations}
Here we present the background and perturbation equations. The background equations are derived upon varying the action with respect to the metric tensor and scalar field. The gravitational equations are given in index notation by:
\begin{align}
    &G_{\mu\nu}-\nabla_\mu\phi\nabla_\nu\phi-\frac{1}{2}g_{\mu\nu}(\nabla\phi)^2\nonumber\\
    &+\frac{\alpha}{g}g_{\mu(\rho}g_{\sigma)\nu}\epsilon^{\kappa\rho\alpha\beta}\epsilon^{\sigma\gamma\lambda\tau}{R}_{\lambda\tau\alpha\beta}\nabla_{\gamma}\nabla_{\kappa}\phi\nonumber\\
    &-\gamma\big[2\nabla_{\nu }\nabla_{\mu }\phi \,\Box\phi-{R} \nabla_{\mu }\phi \nabla_{\nu }\phi - 2 G_{\nu \rho } \nabla_{\mu }\phi \nabla^{\rho } \phi\nonumber\\
    &-2 G_{\mu \rho } \nabla_{\nu }\phi \nabla^{\rho }\phi+ {R}_{\mu \nu } (\nabla_{\mu }\phi)^2  \label{eq:bg_grav} \\
    &- 2\nabla_{\rho }\nabla_{\nu }\phi \nabla^{\rho }\nabla_{\mu }\phi- g_{\mu \nu } (\Box\phi)^2+ G_{\rho \sigma } g_{\mu \nu } \nabla^{\rho 
    }\phi \nabla^{\sigma }\phi\nonumber\\
    &+ g_{\mu \nu } {R}_{\rho \sigma } \nabla^{\rho }\phi \nabla^{\sigma }\phi- 2{R}_{\mu \rho \nu \sigma } \nabla^{\rho }\phi \nabla^{\sigma 
    }\phi + g_{\mu \nu } (\nabla_{\sigma }\nabla_{\rho }\phi)^2\big]\nonumber\\
    &-\sigma\big[\nabla_{\mu}\phi \nabla_{\nu}\phi \;\Box\phi - 2\nabla_{\rho }\nabla_{(\mu}\phi\nabla_{\nu)}\phi \nabla^{\rho }\phi\nonumber\\
    &+ g_{\mu\nu} \nabla^{\rho }\phi \nabla_{\sigma }\nabla_{\rho }\phi \nabla^{\sigma }\phi\big]+\kappa\big[g_{\mu\nu}(\nabla\phi)^2(\nabla\phi)^2/4\nonumber\\
    &-(\nabla\phi)^2\nabla_\mu\phi\nabla_\nu\phi\big]=0\,,\nonumber
\end{align}
while the scalar one is given by:
\begin{align}
    &\Box\phi+\alpha \GB+2\gamma G^{\mu \nu } \nabla_{\nu }\nabla_{\mu }\phi -\kappa\big[(\nabla\phi)^2\Box\phi\nonumber\\
    &+2\nabla^\mu\phi\nabla^\nu\phi\nabla_{\mu}\nabla_{\nu}\phi\big]
    -\sigma \big[\nabla^{\mu }\phi \,\Box\,\nabla_{\mu }\phi\label{eq:bg_scalar}\\
    &+ (\nabla_{\mu}\nabla_{\nu }\phi)^2-\nabla_{\mu }\,\Box\phi \nabla^{\mu}\phi - (\Box\phi)^2\big]=0\,. \nonumber
\end{align}

For the metric introduced in \eqref{eq:metric} the relevant gravitational and scalar background equations are the following:
\begin{align}
    (tt):\quad
    &
    2 B (-2 (8 \alpha  \phi ''+1)+B' \phi ' (24 \alpha +r^2 \sigma  \phi '^2\nonumber\\
    &
    -6 \gamma  r \phi ')-((2 \gamma +r^2) (\phi ')^2))-4 B' (4 \alpha  \phi '\\
    &
    +r)+B^2 (32 \alpha  \phi ''-4 \phi '^2 (\gamma -r^2 \sigma  \phi '')\nonumber\\
    &
    +\kappa  r^2 \phi '^4-16 \gamma  r \phi ' \phi '')+4=0\, ,\nonumber\\[2mm]
    (rr):\quad
    &
    2 A' B (r^2 \sigma  B \phi '^3+8 \alpha  (3 B-1) \phi '-6 \gamma  r B \phi '^2\nonumber\\
    &
    -2 r)+A (B^2 \phi '^2 (-12 \gamma -3 \kappa  r^2 \phi '^2)\nonumber\\
    &
    +8 r \sigma  \phi '+2 B ((2 \gamma +r^2) \phi '^2-2)+4)=0\, ,\nonumber\\[2mm]
    (\phi):\quad
    &
    2 A (\phi ' (4 B r \left(-B \gamma  A''+2 A B \sigma  \phi ''+A\right)\nonumber\\
    &
    +A B' (-6 B \gamma +2 \gamma +r^2))+B \phi '^2 (B (r^2 \sigma  A''\nonumber\\
    &
    -6 A \kappa  r^2 \phi ''+4 A \sigma )+6 A r \sigma  B')\nonumber\\
    &
    +2 B (4 \alpha  (B-1) A''+A \phi '' (-2 B \gamma +2 \gamma\nonumber\\
    &
    +r^2))-A B \kappa  r \phi '^3 (3 r B'+4 B))\nonumber\\
    &
    +A A' (B' (8 \alpha  (3 B-1)+3 B r^2 \sigma  \phi '^2\nonumber\\
    &-12 B \gamma  r \phi ')+2 B (\phi ' (-6 B \gamma +2 B r^2 \sigma  \phi ''\nonumber\\
    &
    +2 \gamma +r^2)-B \kappa  r^2 \phi '^3-4 B \gamma  r \phi
    ''\nonumber\\
    &
    +4 B r \sigma  \phi '^2))+B A'^2 (-8 \alpha  (B-1)\nonumber\\
    &
    -B r^2 \sigma  \phi '^2+4 B \gamma  r \phi ')=0\, .\nonumber
\end{align}
where a prime denotes differentiation with respect to the radial coordinate $r$.

\section{Perturbation equations}
Here, we present the form of the perturbation equations in the frequency domain. These are the equations that we integrate in our search for the QNMs. In the following equations the functions $\mathcal{F}_i$ depend on the background solutions and on the frequency $\omega$. They are rather complicated so we avoid providing their exact form here, but we denote the parameters of the theory upon which they depend.

For the axial QNMs we make use of two independent first order differential equations for the functions $h_0,\,h_1$. The equations have the following form:
\begin{align}
    (r\varphi):\quad
    &
    h_0'-\frac{2 h_0}{r}+\mathcal{F}_1(\alpha,\gamma,\sigma,\kappa)\,h_1=0\, ,\\[2mm]
    (\theta\varphi):\quad
    &
    h_1'+\mathcal{F}_2(\alpha,\gamma)\,h_1+\mathcal{F}_3(\alpha,\gamma)\,h_0=0\, ,
\end{align}
It is straightforward to transform the system of the two independent first order axial differential equations to a single second order one and define the axial potential accordingly. For our numerical integrations it is preferable to work with the system of the two first order differential equations.

For the polar modes, the four independent equations which we use are the following:
\begin{widetext}
\begin{align}
    (tr):\quad
    &
    K'+\mathcal{F}_4(\alpha,\gamma,\sigma)\,H_0+(r^{-1}-A'A^{-1})\,K +\mathcal{F}_5(\alpha,\gamma,\sigma,\kappa)\,H_0+\mathcal{F}_6(\alpha,\gamma,\sigma,\kappa)\,\phi_1+\mathcal{F}_7(\alpha,\gamma,\sigma)\,\phi_1'=0\, ,\\[2mm]
    (t\theta):\quad
    &
    H_1'+\mathcal{F}_8(\alpha,\gamma)\,H_1+\mathcal{F}_9(\alpha,\gamma)\,H_0+\mathcal{F}_9(\alpha,\gamma,)\,K+\mathcal{F}_{10}(\alpha,\gamma)\,\phi_1=0\, ,\\[2mm]
    (r\theta):\quad
    &
    H_0'-\omega A^{-1}H_1+\mathcal{F}_{11}(\alpha,\gamma,\sigma)\,H_0+\mathcal{F}_{12}(\alpha,\gamma)\,K'+\mathcal{F}_{13}(\alpha,\gamma,\sigma,\kappa)\,\phi_1 +\mathcal{F}_{14}(\alpha,\gamma,\sigma)\,\phi_1'=0\, ,\\[2mm]
    (\phi_1):\quad
    &
    \phi_1''+\mathcal{F}_{15}(\alpha,\gamma,\sigma,\kappa)\,H_1+\mathcal{F}_{16}(\alpha,\gamma,\sigma,\kappa)\,H_0' +\mathcal{F}_{17}(\alpha,\gamma,\sigma,\kappa)\,H_0+\mathcal{F}_{18}(\alpha,\gamma,\sigma,\kappa)\, K'\\
    &
    +\mathcal{F}_{19}(\alpha,\gamma,\sigma,\kappa)\, K +\mathcal{F}_{20}(\alpha,\gamma,\sigma,\kappa)\,\phi_1'=0\, . \nonumber
\end{align}
\end{widetext}

\bibliography{bibnote}

\end{document}